\documentclass[twocolumn,secnumarabic,showpacs,showkeys,
amssymb,amsmath,nobibnotes,aps,prl]{revtex4}

\usepackage{graphicx}% Include figure files

\begin{document}

\title{Why the Maxwellian Distribution is the 
Attractive Fixed Point of the Boltzmann Equation}

\author{Ricardo L\'opez-Ruiz}%
\email{rilopez@unizar.es}
%\homepage{http://add.unizar.es/public/100_16613/index.html}
\affiliation{
DIIS and BIFI, Universidad de Zaragoza,\\
E-50009 Zaragoza, Spain.}

\author{Xavier Calbet}
\email{xcalbet@yahoo.es}
\affiliation{
Instituto de Astrof{\'\i}sica de Canarias, V{\'\i}a L\'actea, s/n,\\
E-38200 La Laguna, Tenerife, Spain.\\}

\date{\today}

\begin{abstract}
The origin of the Boltzmann factor is revisited.
An alternative derivation from the microcanonical picture
is given. The Maxwellian distribution in a mono-dimensional ideal gas is 
obtained by following this insight. Other possible applications, as 
for instance the obtaining of the wealth distribution in the human 
society, are suggested in the remarks.  

\end{abstract}

\pacs{02.50.-r, 05.20.-y, 89.65.Gh}
\keywords{Boltzmann factor, Maxwellian distribution, wealth distribution}

\maketitle

We know that the velocity distribution of a gas of classical particles
in equilibrium is the Maxwellian distribution. 
This is a very well experimentally confirmed fact.
The approach in kinetic theory that gives the time evolution of the velocity distribution
of a gas of particles is the Boltzmann equation. 
Hence, the Boltzmann equation should have the Maxwellian distribution as an attractive 
fixed point, i.e., when the initial conditions are far from the equilibrium,
the distribution function, whose time evolution is given by the Boltzmann 
equation, should relax to the Maxwellian distribution. 
This happens when the possibility of binary collisions among 
the particles is considered. This is called the H-Theorem.
Let us remark that this striking result is a non a priori result
in the sense that it is obtained as an approximation to the real
dynamics after ignoring higher order collisions among the particles of the gas.
Thus, the symmetries imposed by the two-body collisions in the asymptotic equilibrium 
state determines correctly the final distribution, i.e., the Maxwellian distribution.

Here we want to offer an alternative view that can be thought as an a priori theoretical argument
explaining why the Maxwellian distribution is the velocity distribution found in a gas 
of classical particles in equilibrium. Usually this fact is shown from the canonical distribution,
that is, the gas is considered in thermal equilibrium with a heat reservoir.
But we do not need the concept of temperature to reach the Maxwellian distribution
nor the supposition of binary collisions in the Boltzmann equation.
It can be easily obtained from the microcanonical picture. Let us proceed to 
show it.

We start by supposing a general one-dimensional ideal gas of $n$ nonidentical 
classical particles with masses $m_i$, with $i=1,\ldots,n$, and with total 
energy $E$. If the particle
$i$ has a momentum $\bar p_i$, we redefine its kinetic energy as
\begin{equation}
p_i^2 = {1 \over 2}{\bar p_i^2 \over m_i}.
\label{eq-p_i}
\end{equation}   
If the total energy is also redefined as $E=R^2$, we have
\begin{equation}
p_1^2+p_2^2+\cdots +p_{n-1}^2+p_n^2 = R^2.
\label{eq-E}
\end{equation}   
We see that the isolated system evolves on the surface of an $n$-sphere. 
Let us recall at this point the formula to calculate the surface $S_n(R)$
of an $n$-sphere of radius $R$:
\begin{equation}
S_n(R) = {2\pi^{n\over 2}\over \Gamma({n\over 2})}R^{n-1},
\label{eq-S_n}
\end{equation}
where $\Gamma(\cdot)$ is the gamma function. If the ergodic hypothesis is supposed, 
each state of the microcanonical ensemble, i.e., each point
on the $n$-sphere, is equiprobable. Then the probability $f(p_n)dp_n$ of finding 
the particle $n$ with coordinate $p_n$ (energy $p_n^2$) is proportional to the 
surface formed by all the points on the $n$-sphere having the $nth$-coordinate 
equal to $p_n$. It is our objective to show that $f(p_n)$ is the Maxwellian 
distribution, with normalization condition
\begin{equation}
 \int_{-R}^Rf(p_n)dp_n = 1.
\label{eq-p_n}
\end{equation}

If the $nth$ particle has coordinate $p_n$, the $(n-1)$ remaining particles 
share the energy $R^2-p_n^2$ on the $(n-1)$-sphere
\begin{equation}
p_1^2+p_2^2+\cdots +p_{n-1}^2 = R^2-p_n^2,
\label{eq-E1}
\end{equation}   
whose surface is $S_{n-1}(\sqrt{R^2-p_n^2})$. 
If we define the coordinate $\theta$ as following:
\begin{equation}
 R^2\cos^2\theta = R^2-p_n^2,
\label{eq-theta}
\end{equation}
then
\begin{equation}
 Rd\theta = {dp_n \over (1-{p_n^2\over R^2})^{1\over 2}}.
\label{eq-diftheta}
\end{equation}
It can be easily proved that
\begin{equation}
 S_n(R) = \int_{-{\pi\over 2}}^{\pi\over 2} S_{n-1}(R\cos\theta)Rd\theta.
\label{eq-theta1}
\end{equation}
Rewriting this last expression as function of $p_n$ we obtain: 
\begin{equation}
 {1\over S_n(R)}\int_{-R}^R
 S_{n-1}(\sqrt{R^2-p_n^2}){dp_n \over (1-{p_n^2\over R^2})^{1\over 2}} = 1,
\label{eq-p_n1}
\end{equation}
and comparing with condition (\ref{eq-p_n}), the expression for $f(p_n)$
is obtained:
\begin{equation}
 f(p_n) = {1\over S_n(R)}
{ S_{n-1}(\sqrt{R^2-p_n^2})\over (1-{p_n^2\over R^2})^{1\over 2}},
\label{eq-f_n}
\end{equation}
whose final form, after some calculations, results to be
\begin{equation}
f(p_n) = C_n R^{-1}\left(1-{p_n^2\over R^2}\right)^{n-3\over 2},
\label{eq-mm}
\end{equation}
with
\begin{equation}
C_n = {1\over\sqrt{\pi}}{\Gamma({n\over 2})\over \Gamma({n-1\over 2})}.
\label{eq-cn}
\end{equation}

If we call $\epsilon$ the mean energy per particle, $E=R^2=n\epsilon$,
we have for $p_n^2\ll R^2$:
\begin{equation}
\left(1-{p_n^2\over R^2}\right)^{n-3\over 2}\simeq e^{-{p_n^2\over 2\epsilon}{n-3\over n}}.
\label{eq-ee}
\end{equation}
The Boltzmann factor $e^{-{p_n^2\over 2\epsilon}}$ is found 
when ${n-3\over n}\simeq 1$, and this happens
with an error less than $10\%$ when $n>30$. This means that the behavior of a 
gas with $30$ particles or more presents a similar statistical behavior for
the low energy particles than a many particle gas. When $n\gg 1$, 
the Stirling approximation can be applied to expression (\ref{eq-cn}), then we have:
\begin{equation}
\lim_{n\gg 1} C_n \simeq  {1\over\sqrt{\pi}}\sqrt{n\over 2},
\label{eq-cc}
\end{equation}
and substituting expressions (\ref{eq-ee})-(\ref{eq-cc})
in (\ref{eq-mm}) the expression of the Maxwellian distribution is 
obtained in the asymptotic regime $n\rightarrow\infty$ 
(this also implies $E\rightarrow\infty$):
\begin{equation}
f(p)dp =  \sqrt{1\over 2\pi\epsilon}\;\;e^{-{p^2\over 2\epsilon}}dp,
\end{equation}
where the index $n$ has been already removed. This indicates the independence
of this function on the behavior of a particular particle and, in consequence, 
the possibility of calculating the velocity distribution averaging
over all the gas. 

Depending on the physical situation the mean energy per particle $\epsilon$
takes different expressions. For a mono-dimensional gas in thermal equilibrium, 
the theorem of equipartition of energy implies that $\epsilon={1\over 2}kT$, 
with $k$ the Boltzmann constant and $T$ the temperature of
the gas. If $p^2$ is replaced by ${1\over 2}mv^2$, the Maxwellian 
distribution as function of particle velocity, and as it is usually presented
in the literature, is finally recovered:
\begin{equation}
g(v)dv =  \sqrt{m\over 2\pi kT}\;\;e^{-{mv^2\over 2kT}}dv.
\end{equation}

$\;$\par \vspace{0.1cm}

{\bf Remark 1:} The origin of this note is in the manuscript \cite{calbet06},
where the same calculation was performed in spherical coordinates 
(equation (6) of that paper).

{\bf Remark 2:} When we were writing this note we have been aware 
of an article \cite{stam}, published in the mathematical 
context of statistics, where this result 
is presented in a more general form. The abstract of this paper says:
``If $X=(X_1,X_2,\ldots,X_n)$ has uniform distribution on the sphere or ball
in $\Re^n$ with radius $a$, then the joint distribution of $n^{1\over 2}a^{-1}X_i$,
$i=1,2,\ldots,k$, converges in total variation to the standard normal 
distribution on $\Re^k$''. The case here shown is $k=1$.  

{\bf Remark 3:} This calculation shows that the Boltzmann factor 
is the general statistical behavior of each small part 
of a multi-component system whose components or parts are given
by a set of random variables that verify some conservation law (condition (\ref{eq-E})).
For instance, the derivation of this factor in the context of the canonical
ensemble could also be performed by following this insight. A more interesting 
application can be to obtain the distribution of wealth in the human society.
It is known that the incomes of the $90\%$ of the population in western societes
can be fitted by a Gibbs exponential distribution \cite{yakovenko}, and then, 
this experimental confirmation can receive a simple explanation if we interpret 
that, at each time, a fixed quantity of wealth or money (condition (\ref{eq-E}))
is randomly partitioned among many agents.

\end{document}